\documentclass{natureprintstylev4}
\usepackage{astjnlabbrev-nature}
\usepackage[switch]{lineno}

\usepackage{ulem}
\usepackage{array}
\usepackage{amssymb}
\usepackage{mathptmx}
\usepackage{amsmath}	
\usepackage{gensymb}
\usepackage{color}
\usepackage{enumitem}
\usepackage{textcomp}
\usepackage{marvosym}

\usepackage{epsfig}
\usepackage{color}
\usepackage{graphicx}
\usepackage{longtable}
\usepackage{float}
\usepackage{orcidlink}
\usepackage{hyperref}
\hypersetup{
    colorlinks=true,
    linkcolor=blue,
    filecolor=blue,      
    urlcolor=blue,
    citecolor=blue,
}

\usepackage[switch]{lineno}

\usepackage[labelsep=endash]{caption}

\newcommand{\citep}{\cite}
\newcommand{\citet}{\cite}
\usepackage[utf8]{inputenc}

\title{Comment on `a slightly oblate dark matter halo revealed by a retrograde precessing Galactic disk warp' by Huang et al.}

\author{Walter Dehnen\,\orcidlink{0000-0001-8669-2316}%
$^{1\textsuperscript{\Letter}}$,
Ralph Sch{\"o}nrich$^2$,
Ronald Drimmel\,\orcidlink{0000-0002-1777-5502}%
$^3$,
\v{Z}ofia Chrob\'akov\'a\,\orcidlink{0000-0002-9895-6638}%
$^2$,
Eloisa Poggio\,\orcidlink{0000-0003-3793-8505}%
$^{3}$ and \\
Marcin Semczuk\,\orcidlink{0000-0002-8191-8918}%
$^{4,5,6}$}

\begin{document}
\maketitle
\begin{affiliations}
\item Astronomisches Rechen-Institut, Zentrum f{\"u}r Astronomie der Universit{\"a}t Heidelberg, M{\"o}nchhofstr.~12-14, 69120 Heidelberg, Germany;\\
\item Mullard Space Science Laboratory, University College London, Holmbury St. Mary, RH56NT Dorking, United Kingdom;\\
\item INAF - Osservatorio Astrofisico di Torino, via Osservatorio 20, 10025 Pino Torinese (TO), Italy;\\
\item Departament de Física Quàntica i Astrofísica (FQA), Universitat de Barcelona (UB), C Martí i Franqués, 1, 08028 Barcelona, Spain;\\
\item Institut de Ciències del Cosmos (ICCUB), Universitat de Barcelona (UB), C Martí i Franqués, 1, 08028 Barcelona, Spain;\\
\item Institut d'Estudis Espacials de Catalunya (IEEC), Edifici RDIT, Campus UPC, 08860 Castelldefels (Barcelona), Spain;\\
$\textsuperscript{\Letter}$ Corresponding author: walter.dehnen@uni-heidelberg.de.
\end{affiliations}

\section*{Abstract}
\textbf{
Huang et al.~(2024)\cite{Huang2024} measured the derivative of the phase $\phi_{\mathrm{w}}$ of the Galactic warp traced by classical Cepheids with respect to their age $\tau$ and interpreted it as the warp precession rate $\omega\equiv\mathrm{d}\phi_{\mathrm{w}}/\mathrm{d}t=-\mathrm{d}\phi_{\mathrm{w}}/\mathrm{d}\tau$. This interpretation is unfounded: young stars follow trajectories close to those of their parental gas and trace the instantaneous gas warp, not its shape at their time of birth: $\phi_{\mathrm{w}}$ should hardly depend on Cepheid age. We show that the measured $\mathrm{d}\phi_{\mathrm{w}}/\mathrm{d}\tau>0$ is consistent with an omitted-variable bias from neglecting the natural twist $\mathrm{d}\phi_{\mathrm{w}}/\mathrm{d}R$ of the warp and the $R$-$\tau$ correlation for Cepheids (originating from the Galactic metallicity gradient and the Cepheid metallicity-age correlation).}

For most disc galaxies with orientation amenable to warp detection, a warp is observed\cite{SanchezSaavedraEtAl1990, Bosma1991, GarciaRuiz2002}. Applying a tilted-ring model to galactic H\,{\footnotesize I} discs, Briggs\cite{Briggs1990} found their outer parts to be warped with typical inclinations of $i=5^\circ$-20$^\circ$. For small inclinations, the tilted rings have vertical displacement
\begin{align}
    \label{eq:warp:z}
    z &= Z \sin(\phi-\phi_{\mathrm{w}}),
\end{align}
with warp amplitude $Z=R\tan i$ and phase $\phi_{\mathrm{w}}$, which equals the azimuth of the ascending node. Typically, $\phi_{\mathrm{w}}$ increases outward in the direction of galactic rotation and forms a leading spiral, known as `Briggs' rule'. The Milky-Way warp was detected in radio emission from H\,{\footnotesize I}\citep{Burke57, Westerhout1957, Henderson1982, Levine2006}, but the precise shape and run of $\phi_{\mathrm{w}}(R)$ cannot be measured from H\,{\footnotesize I} since the distances to the radio emitters are not known to sufficient accuracy. More recently, the Milky-Way warp has also been traced by classical Cepheids and found to be well described by equation~\eqref{eq:warp:z}\citep{Chen2019, Skowron2019A, Skowron2019B, Lemasle2022, Dehnen2023, Cabrera-Gadea2024}. In a simplistic warp model the gravitational potential is assumed static and the warp precesses at each radius like an inclined orbit with rate\citep{Binney1992}
\begin{align}
    \label{eq:om:p}
    \omega \equiv \mathrm{d}\phi_{\mathrm{w}}/\mathrm{d}t = \Omega_\phi-\Omega_z.
\end{align}
Here, $\Omega_\phi$ and $\Omega_z$ are the frequencies of rotation and vertical oscillation, respectively, which depend on the gravitational potential. In general $\Omega_\phi\simeq\Omega_z$ and therefore $|\omega|\ll\Omega_\phi$: warp precession is slow. For oblate mass distributions, $\Omega_z>\Omega_\phi$ such that $\omega<0$: warp precession is retrograde. Moreover, the orbital frequencies decline with $R$: the warp precesses faster at smaller radius and with time its line of nodes winds into a leading spiral (for retrograde precession), conforming to Briggs's rule. This simple model can approximate the long-term mean of $\omega$, but in the short-term warp precession is altered by spiral patterns\cite{MassetTagger1997} and satellite impacts\cite{Weinberg1998,Poggio2021}, so that the instantaneous $\omega$ cannot constrain the Galactic potential.

\begin{figure}
    \centering
    \includegraphics[width=\columnwidth]{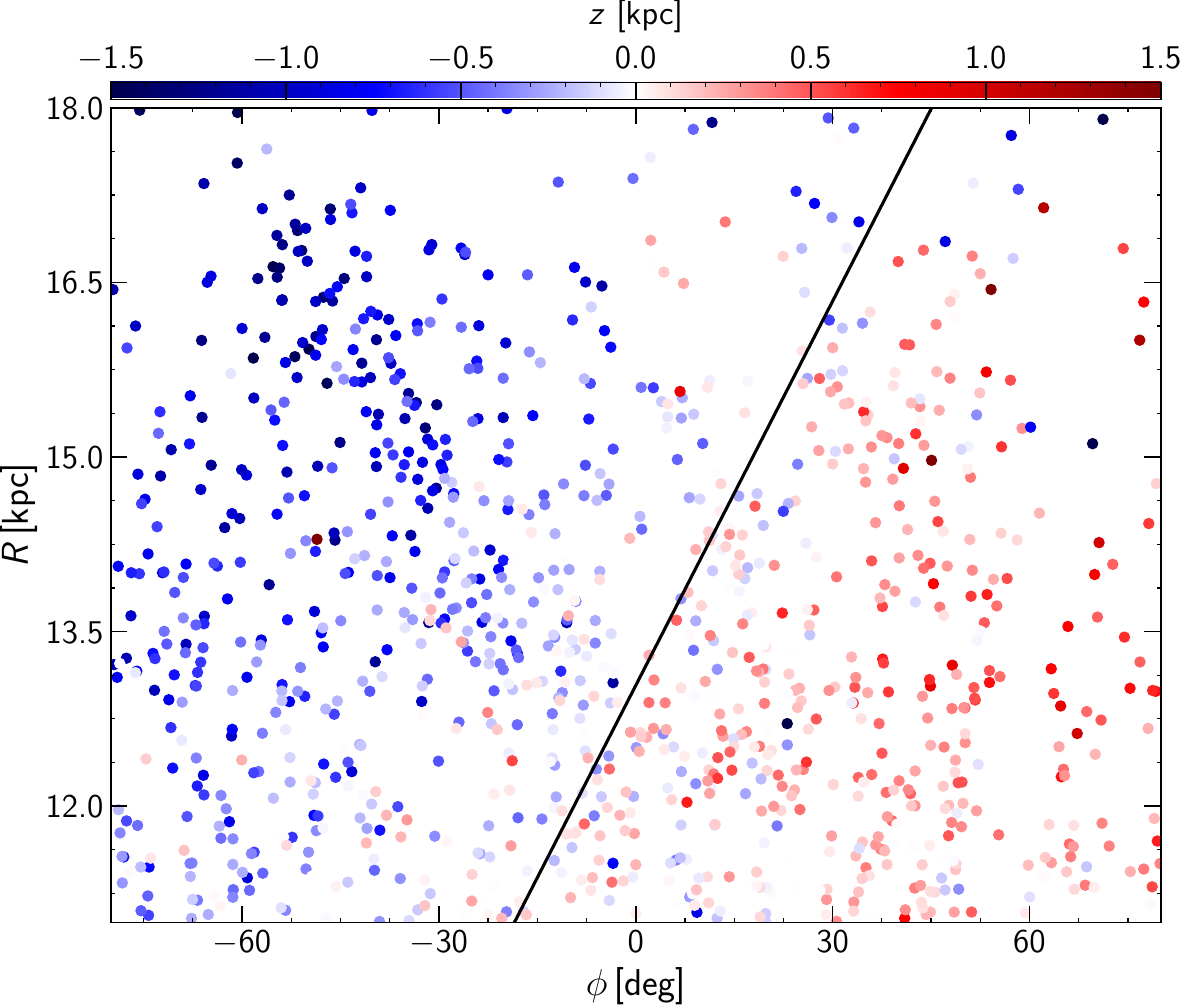}
    \caption{Distribution of Milky-Way Cepheids\cite{Skowron2024} over Galactic radius $R$, azimuth $\phi$ (increasing in direction of rotation), and $z$ (colour) in the part of the outer disc containing the ascending nodes ($z=0$) of the warp. These are well described by the plotted line\cite{Dehnen2023}, which maps to a leading spiral in ($x,y$).}
    \label{fig:R:phi}
\end{figure}

\begin{figure}
    \centering
    \includegraphics[width=\columnwidth]{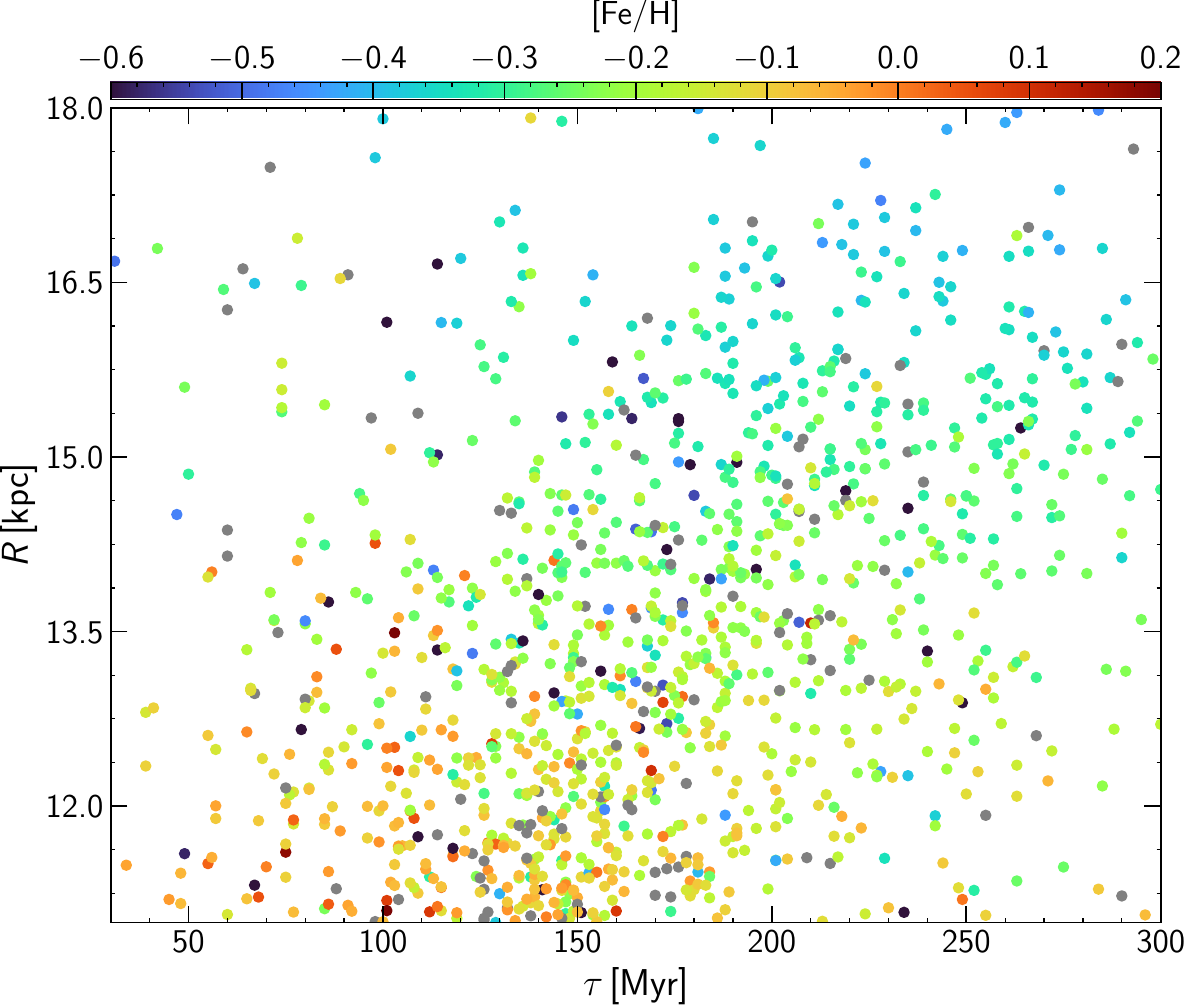}
    \caption{Distribution of the Milky-Way Cepheids\cite{Skowron2024} over Galactic radius $R$, age$^3$ $\tau$, and [Fe/H] (indicated by colour, grey for stars without information). Pearson's correlation coefficient for the stars in this plot is $\rho_{R\tau}=0.35$.}
    \label{fig:R:age}
\end{figure}

Nonetheless, with this very aim Huang et al.\cite{Huang2024} (hereafter H24) attempted to measure $\omega$ from Cepheids. To this end, they employ a `motion-picture' method, which uses stellar age as look back time and gives $\omega=-\mathrm{d}\phi_{\mathrm{w}}/\mathrm{d}\tau$. Unfortunately, this does not work: Cepheids are subject to the same forces as their surroundings, follow almost the same trajectories as their parental gas, and partake in the instantaneous warp, just like other stars and gas, regardless of their age, i.e.\ we expect $\mathrm{d}\phi_{\mathrm{w}}/\mathrm{d}\tau\simeq0$. Nonetheless, H24 (see their Fig.~1d) measured 
\begin{align}
    \label{eq:dpsi/dtau}
    \mathrm{d}\phi_{\mathrm{w}}/\mathrm{d}\tau = (0.121\pm0.03)\,\mathrm{deg/Myr}.
\end{align}
This seemingly unexpected result is completely spurious and can be understood in terms of an effect known as `omitted-variable bias'. This requires a neglected relation of $\phi_{\mathrm{w}}$ with a variable that correlates with Cepheid age. This variable is the Galactic radius $R$, since for galaxies following Briggs' rule, including the Milky Way, the warp phase $\phi_{\mathrm{w}}$ increases with $R$. We demonstrate this in Fig.~\ref{fig:R:phi}, which plots the distribution over $R$, $\phi$, and $z$ of Galactic Cepheids in the region containing the line of nodes of the warp. The slanted line indicates the twist\cite{Dehnen2023}
\begin{align}
    \label{eq:dpsi/dR:D23}
    \mathrm{d}\phi_{\mathrm{w}}/\mathrm{d}R = (10.6\pm0.8)\,\mathrm{deg/kpc}
\end{align}
of the warp line of nodes, determined from essentially the same set of stars (using the Cepheid's guiding-centre radius $R_\mathrm{g}$ instead of direct radius $R$, but this difference does not matter here). Galactic radius $R$ correlates with Cepheid age $\tau$\cite{Skowron2019B, Anders2024}, as we demonstrate in Fig.~\ref{fig:R:age}, because of the radial Galactic metallicity gradient and since metal-poor Cepheids are on average older than their metal-rich counterparts. This in turn is because with declining metallicity (1) Cepheid progenitors of a given initial mass take longer to reach the instability strip, and (2) the lowest initial-mass of stars evolving to become Cepheids decreases (from 4.5\,M$_\odot$ to 3\,M$_\odot$ at Solar and 0.25 Solar metallicity, respectively\cite{Anderson2014, Anderson2016}).

In conjunction with the radial gradient~\eqref{eq:dpsi/dR:D23} of $\phi_{\mathrm{w}}$, this correlation between $R$ and $\tau$ mimics an age-dependence of $\phi_{\mathrm{w}}$. Quantitatively, fitting the linear relation $\phi_{\mathrm{w}}=\alpha_\tau+\beta_\tau\tau$ to the data gives for the slope
\begin{align}
    \label{eq:beta:fit}
    \beta_\tau=\frac{\langle\phi_{\mathrm{w}}\tau\rangle-\langle\phi_{\mathrm{w}}\rangle\langle\tau\rangle}{\langle\tau^2\rangle-\langle\tau\rangle^2},
\end{align}
where $\langle\cdot\rangle$ denotes a sample average over the Cepheids. Inserting the true linear relation $\phi_{\mathrm{w}}=\alpha_R + \beta_R R$ obtains
\begin{align}
    \label{eq:OVB}
    \beta_\tau = \beta_R\,\rho_{R\tau}\,(\sigma_R/\sigma_\tau),
\end{align}
where $\rho_{R\tau}$ is Pearson's correlation coefficient and $\sigma$ the standard deviations. We measured these quantities for Cepheids at $R>11\,$kpc in the latest catalogue of 3358 Galactic Cepheids\cite{Skowron2024} (ages are determined via a period-age-metallicity relation\cite{Anderson2016}, different from that\cite{DeSomma2021} used by H24), obtaining $\beta_\tau \sim 0.01\,\beta_R\, \mathrm{kpc/Myr}$. With the known radial gradient $\beta_R$ (equation~\ref{eq:dpsi/dR:D23}), this gives $\beta_\tau \sim 0.1\,\mathrm{deg/Myr}$, consistent with the value~\eqref{eq:dpsi/dtau} measured by H24.

Despite minor differences in the analysis (age determination, sample), this demonstrates that $\mathrm{d}\phi_{\mathrm{w}}/\mathrm{d}\tau$ measured by H24 is fully consistent with being an omitted-variable bias from neglecting the known twist~\eqref{eq:dpsi/dR:D23} of the Galactic warp (Fig.~\ref{fig:R:phi}) and the Cepheid radius-metallicity correlation (Fig.~\ref{fig:R:age}). By ignoring these relations, H24 measured a spurious derivative $\mathrm{d}\phi_{\mathrm{w}}/\mathrm{d}\tau$, which together with an erroneous equating of Cepheid age to look-back time resulted in a seemingly sensible value for $\omega$, even though it deviates from all previous measurements.

H24 also attempted to measure the warp precession rate kinematically, from the rate of change of the fitted warp phase implied by the stellar velocities. Their result $\omega=(-1.1\pm1.9)\,$Gyr$^{-1}$ deviates substantially from $\omega\sim10\,$Gyr$^{-1}$ obtained previously with similar methods and essentially the same set of stars\cite{Poggio2020, Cheng2020, Dehnen2023, Cabrera-Gadea2024}.

We repeated the kinematic measurement of $\omega$ from the latest Cepheid catalogue\cite{Skowron2024} and obtained values consistent with previous measurements, unless we follow H24 and apply a vertical velocity offset of $\Delta v_z=4.16\,$km\,s$^{-1}$ to all stars. H24 argue that this offset ``\textit{correct[s] for the possible non-zero vertical velocity at the starting radius [of the warp]}''. A velocity offset of this magnitude is highly unexpected, but should be measured as $m=0$ azimuthal Fourier component independently of the warp ($m=1$ component), rather than being `corrected' upfront.

We conclude that H24 did not in fact measure the warp precession rate $\omega$ and hence that any conclusion based on their value for $\omega$ is void.

\section*{Data Availability}
Cepheid data used were from the latest catalogue of Galactic Cepheids\cite{Skowron2024}.

\section*{Code Availability}
We use standard publicly available Python modules astropy, numpy, and matplotlib.

\section*{Acknowledgements}
We thank Shourya Khanna for help with the data preparation. 
RS and ZC acknowledge funding from the Royal Society. 

\section*{Contributions}
WD instigated the project, performed the analysis, and wrote the text; RS, RD, ZC, EP, and MS contributed to the analysis and/or the text.

\section*{Competing interests}
The authors declare no competing interests.

\section*{References}
\bibliography{refs}

\end{document}